\documentclass[twoside]{article}
\usepackage{fleqn,espcrc2}
\usepackage{graphicx,epsfig}

\title{Exclusive hadronization in
  $\gamma^*\gamma\to\pi\pi$~\thanks{Talk given at the International
  Conference on the Structure and Interactions of the Photon (PHOTON
  99), Freiburg im Breisgau, Germany, 23--27 May 1999, to appear in
  Nucl.\ Phys.\ B (Proc.\ Suppl.)}}

\author{M. Diehl\address{Deutsches Elekronen-Synchroton DESY, 22603
Hamburg, Germany}, T. Gousset\address{Subatech, B.P. 20722, 44307
Nantes, France} and B. Pire\address{Centre de Physique Th\'eorique,
\'Ecole Polytechnique, 91128 Palaiseau, France}.}

\begin{document}

\begin{abstract}
  We perform a QCD analysis of exclusive two-meson production in
  $\gamma^* \gamma $-collisions in the kinematical domain of large
  photon virtuality $Q^2$ and small hadronic invariant mass $W^2$. In
  these kinematics the amplitude factorizes into a perturbative
  subprocess $\gamma^* \gamma \to q \bar q$ and a two-meson
  distribution amplitude.  This opens the way to study new hadronic
  matrix elements related with the physics of hadronization and
  confinement.  \vspace{1pc}
\end{abstract}

\maketitle

\section{INTRODUCTION}

The physics of the reaction $\gamma^* \gamma \to \pi \pi$ in the
kinematical domain where the photon virtuality is large compared to
the hadronic invariant mass has recently been shown to be very
interesting from a QCD point of view (\cite{DGPT}, see also the
pioneering work by D.  M\"uller {\it et al.}~\cite{MRG}). The reason
is that this process factorizes~\cite{Fre} into a perturbatively
calculable, short-distance dominated subprocess $\gamma^* \gamma \to q
\bar q$ and a non-perturbative QCD matrix element we have called
generalized distribution amplitude (GDA). In fact, this generalizes
the factorization of the $\gamma$--$\pi$ transition form factor
measured in $\gamma^\ast \gamma \to \pi^0$, which has been extensively
studied theoretically and experimentally~\cite{RR,pi-exp}.

Note that $\gamma^* \gamma \to \pi \pi$ is related by crossing to deep
virtual Compton scattering on a meson, $\gamma^\ast \pi \to \gamma
\pi$, which at large $Q^2$ and small squared momentum transfer between
the hadrons factorizes into a hard photon-parton scattering and a
skewed (i.e.\ off-forward, non-forward, or non-diagonal) parton
distribution~\cite{OFPD}. The $\gamma^\ast \gamma$ and Compton
processes share many features, in particular their scaling behavior in
$Q^2$ and a helicity selection rule for the virtual photon.

\section{KINEMATICS}

The reaction we are interested in is either
\begin{equation}
e(k) + e(l) \to e(k') + e(l') + \pi^a(p) + \pi^b(p')
\label{ee}
\end{equation}
or
\begin{equation}
e(k) + \gamma (q') \to e(k')  + \pi^a(p) + \pi^b(p'),
\label{egamma}
\end{equation}
where $a,b = +,0,-$ and
\begin{eqnarray}
q &=& k - k',\ \ \ \ \,Q^2\ =\ -q^2,\nonumber \\
q'&=&l - l',\ \ \ \ \ \ q'^2\ \to\ 0,\nonumber \\
P &=& p+p',\ \ \ \ \hspace{-0.4pt}W^2\ =\ P^2,\nonumber \\
s_{ee}&=& (k+l)^2,\ \ s_{e\gamma}\ =\ (k+q')^2.
\end{eqnarray}

In the $\gamma^* \gamma$ center-of-mass frame, choosing the $z$ axis
along $\mathbf{q}$, natural variables are $\theta$, the polar angle of
$\mathbf{p}$, and $\varphi$, the angle between the leptonic plane and
the hadronic plane. To completely specify the kinematics one needs
three further variables, which can be chosen as $Q^2$, $W^2$ and
$y=(W^2+Q^2)/s_{e\gamma}$.

We shall also use the variable $\zeta$, defined as the light-cone
fraction of momentum carried by $\pi^a(p)$ with respect to the pion
pair. It is related to $\theta$ through
\begin{equation}
\zeta=\frac{1}{2}\left(1+\sqrt{1-\frac{4m_\pi^2}{W^2}}\cos\theta\right).
\end{equation}

\section{FACTORIZATION}

The spacetime cartoon of the process one can derive from power
counting and factorization arguments is shown in Fig.~1.

\begin{figure}
\centerline{\psfig{figure=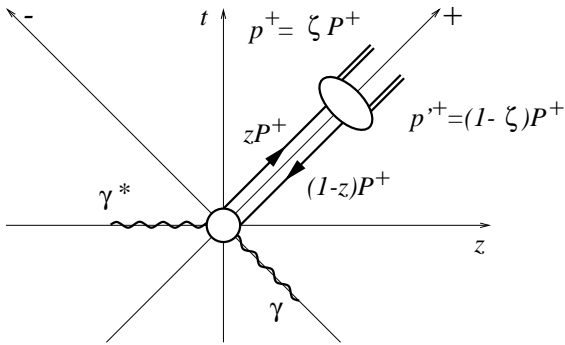,width=\hsize}}
\caption{Spacetime diagram in the Breit frame, obtained by boosting
  from the center of mass along $z$.}
\end{figure}

In the Breit frame the real photon moves fast along $-z$ and is
scattered into an energetic hadronic system moving along $+z$. The
hard part of this process takes place at the level of elementary
constituents, and the minimal number of quarks and gluons compatible
with conservation laws (color etc.) emerge. At Born level one has
simply $\gamma^*\gamma\to q\bar{q}$, but through a quark box the
photons can also couple to two gluons. Each quark or gluon carries a
fraction $z$ or $1-z$ of the large light-cone momentum component
$P^+$. Subsequently the soft part of the reaction, i.e.\ hadronization
into a pion pair, takes place.

At leading order in $\alpha_S$ one has for the hadronic tensor
\begin{eqnarray}
T^{\mu\nu}&=&i\int\! d^4 x\, e^{-i q\cdot x} \,
    \langle \pi \pi |\,
    T J_{\mathrm{em}}^\mu(x) J_{\mathrm{em}}^\nu(0) \,| 0 \rangle
    \nonumber \\
&=&-g^{\mu\nu}_T  %\cdot\nonumber \\
\sum_q \frac{e_q^2}{2}\int_0^1\! dz\,{2z-1\over z(1-z)}\,
\Phi_q,\label{tensor}
\end{eqnarray}
where $g^{\mu\nu}_T$ denotes the metric tensor in transverse space.
Here we introduced for each quark flavor the generalized distribution
amplitude
$$
\Phi_q=\int\frac{dx^-}{2\pi}e^{-iz(P^+ x^-)}\,
\langle\pi\pi|\,\bar{q}(x^-)\gamma^+ q(0)\,|0\rangle,
$$
in light cone gauge $A^+ = 0$. $\Phi_q$ depends on the quark
light-cone fraction $z$, on the kinematical variables $\zeta$ and
$W^2$ of the pions, and on its factorization scale.

Contracting with the photon polarization vectors we see that in order
to give a nonzero $\gamma^\ast \gamma \to \pi \pi$ amplitude the
virtual photon must have the same helicity as the real one, in
particular it must be transverse. As in the case of virtual Compton
scattering this is a direct consequence of chiral invariance in the
collinear hard scattering process~\cite{DGPR}. We also find that the
$\gamma^\ast \gamma$ amplitude is independent of $Q^2$ at fixed
$\zeta$ and $W^2$, up to logarithmic scaling violation to be discussed
later.

There will of course be power corrections in $1/Q$ to this leading
mechanism. An example is the hadronic component of the real photon,
which might be modeled by vector meson dominance. {}From power
counting arguments one obtains that this contribution is suppressed,
as it is in virtual Compton scattering. Our process can also be
treated within the operator product expansion, which allows for a
systematic analysis of higher twist effects.

\section{RELATION WITH THE PHOTON STRUCTURE FUNCTIONS}

The exclusive process we consider here may be seen as the limit of the
inclusive reaction $\gamma^\ast\gamma\to X$ \cite{KSZ}, when unlike in
the Bjorken limit the hadronic mass squared $W^2$ is not scaled up
with $Q^2$ but remains fixed and small enough for the final state $X$
to be saturated by the $\pi\pi$ state. Let us examine the connection
to some aspects of the photon structure functions, pointing out the
specificity of the present kinematical limit.

The differential unpolarized cross section for the reaction
$e+\gamma\to e+X$ is
\begin{eqnarray}
\lefteqn{\frac{d\sigma}{dx_BdQ^2}=\frac{4\pi\alpha^2}{x_BQ^4}}
\nonumber \\ 
&&
\times \bigl\{[1+(1-y)^2]\, x_B F^\gamma_T+(1-y)F^\gamma_L\bigr\},
\end{eqnarray}
where in addition to the kinematical notation of Section 2 we
introduced $x_B=Q^2/(Q^2+W^2)$. $\alpha$ is the QED coupling, and
$F^\gamma_T$ and $F^\gamma_L$ are the photon structure functions, the
index indicating the polarization of the virtual photon. Their
connection to the other familiar set, $F_1$ and $F_2$, is given by
$F_1=F_T$ and $F_2=2x_BF_T+F_L$.

The pointlike contribution of a light quark with charge $e_q$ to
$F^\gamma_T$ and $F^\gamma_L$, computed at zeroth order in the strong
coupling is
\begin{eqnarray}
F^\gamma_T\!&\!=\!&\!\frac{3\alpha}{2\pi}e_q^4 \biggl\{
[x_B^2+(1-x_B)^2]\ln\frac{(1-x_B)Q^2}{x_B m_q^2} \nonumber \\
&&\hspace{3em} -(2x_B-1)^2\biggr\},\\
F^\gamma_L\!&\!=\!&\!\frac{3\alpha}{2\pi}e_q^4
\bigl\{8x_B^2(1-x_B)\bigr\},
\end{eqnarray}
in the Bjorken limit, where $Q^2, W^2 \gg m_q^2$. Note that $m_q$ is
to be understood here as an effective mass that regulates the
divergence in the box diagram. We see that in the region of fixed
$W^2$ with $Q^2\gg W^2$ (where $1-x_B\sim W^2/Q^2$), these
functions are
\begin{eqnarray}
F^\gamma_T&\sim&\frac{3\alpha}{2\pi}e_q^4 \left\{
\ln\frac{W^2}{m_q^2}-1\right\},\\
F^\gamma_L&=&O\left(\frac{W^2}{Q^2}\right),
\end{eqnarray}
which shows the disappearance of the $\ln Q^2$ behavior in
$F^\gamma_T$ and the vanishing of $F^\gamma_L$ as expected.

Let us also stress that the hadronic part of the photon structure
function (usually parameterized by a vector dominance ansatz), which in
general is only suppressed by a factor $\ln Q^2$ with respect to the
pointlike part, does not survive our particular limiting procedure. As
already noted it becomes a $O(1/Q)$ correction.

\section{GENERALIZED DISTRIBUTION AMPLITUDES.}

The physics contained in generalized distribution amplitudes goes
beyond that of a $q\bar{q}$ distribution amplitude of a meson: since
two hadrons are formed $\Phi$ does not select their lowest Fock
states; in this respect it is related to ordinary parton distributions
and to fragmentation functions. If, however, $W$ is at or near the
mass of a resonance with appropriate quantum numbers, such as an
$f_0$, it will contain physics of the distribution amplitude for the
resonance and of its decay into two pions~\cite{Pol}.

The same crossing procedure that connects our $\gamma^* \gamma$
process with deep virtual Compton scattering allows one to relate
generalized distribution amplitudes to skewed parton distributions by
an analytic continuation in the invariant mass variable
$W^2$~\cite{Poly-W}.

Whereas time reversal invariance constrains ordinary distribution
amplitudes and parton distributions to be real valued functions up to
convention dependent phases, our generalized distributions are
\emph{complex}. We notice in (\ref{tensor}) that the hard scattering
kernel at Born level is purely real so that the imaginary part of the
$\gamma^\ast \gamma$ amplitude is due to $\mathrm{Im}\, \Phi$; it
corresponds to rescattering and resonance formation in the soft
transition from the partons to the final state hadron pair. In fact,
Watson's theorem allows to deduce the phase of $\Phi(z,\zeta,W^2)$
from the phase shift analysis of $\pi\pi$ scattering for $W$ not too
large~\cite{Pol}.

Up to now we have discussed generalized $q\bar{q}$ distribution
amplitudes. Beyond tree level in the hard scattering the hadron pair
can however also originate from two gluons. Since gluons are known to
be important in fragmentation and in parton distributions, one can
expect the generalized $gg$ distribution to be of the same order as
the one for $q\bar{q}$.

QCD radiative corrections to the hard scattering will as usual lead to
logarithmic scaling violation. At this point it is useful to remember
the analogy of our process with the transition $\gamma^* \gamma \to
\pi^0$. The evolution can be obtained from the hard scattering kernel
alone and remains the same if we replace $\Phi$ with the distribution
amplitude of a meson with the appropriate quantum numbers, say an
$f_0$. The generalized distribution amplitudes thus follow the usual
Efremov-Radyushkin-Brodsky-Lepage evolution~\cite{ERBL} for a meson.
Notice the peculiarity of the present channel that generalized
$q\bar{q}$ and $g g$ distributions mix under evolution~\cite{Chase}.

\section{PHENOMENOLOGY}

\subsection{The Bremsstrahlung subprocess}

In $e\gamma$ collisions the process $\gamma^\ast \gamma \to h \bar{h}$
we want to study competes with bremsstrahlung, where the hadron pair
$h \bar{h}$ originates from a virtual photon radiated off the
lepton~\cite{Bud}.  This process produces the pair in the $C$-odd
channel and hence does not contribute for $h = \bar{h}$, in particular
not for $h = \pi^0$.  Its amplitude can be computed from the value of
the timelike elastic form factor measured in $e^+ e^- \to h \bar{h}$.
For pions its magnitude and phase are dominated by the $\rho$ meson
peak in a broad mass region around 800 MeV.

\subsection{Interfering subprocesses and C-odd observables}

The $\gamma^\ast \gamma$ and bremsstrahlung processes interfere in the
$\pi^+ \pi^-$ channel; this provides an opportunity to study the
$\gamma^\ast \gamma$ contribution at \emph{amplitude} level. Thanks to
the different charge conjugation properties of the two processes their
interference term can be selected by the charge asymmetries
\begin{equation}
d\sigma( \pi^+(p) \pi^-(p') ) -
d\sigma( \pi^-(p) \pi^+(p') )
\end{equation}
or
\begin{equation}
d\sigma (e^+ \gamma \to e^+ \pi^+ \pi^-) -
d\sigma (e^- \gamma \to e^- \pi^+ \pi^-),
\end{equation}
while it drops out in the corresponding charge averages. We note that
bremsstrahlung has an amplitude with both real and imaginary parts,
especially at values of $W$ where it is dominated by vector meson
resonances, and in particular benefits from the $\rho$ peak at $W$
around 800 MeV.

\subsection{Angular dependence}

The dependence of the $\gamma^* \gamma$ process on $\theta$ and
$\varphi$ is entirely due to the decomposition of the $\pi\pi$ final
state into partial waves, while in the limit $W^2 \ll Q^2$ the
interference contribution has an additional factor $\sin(\theta)
\cos(\varphi)$ coming from the bremsstrahlung amplitude.  This leads
to very specific patterns which will help testing the helicity
selection rules valid in the large-$Q^2$ limit, and thus the
applicability of our theoretical description, at a finite value of
$Q^2$. We will show elsewhere~\cite{Us} these distributions with
specific model assumptions for the generalized distribution
amplitudes.

\section{UNIVERSALITY}

A new hadronic matrix element such as the generalized distribution
amplitude is of great interest since it appears as a universal
quantity in different processes. This is indeed the case as can be
seen from an analysis of the electroproduction process
{$\gamma^*N\to\pi\pi N$}~\cite{Pol,JLAB}.

The amplitude for deep electroproduction of one meson ($\pi, \rho$,
\dots) in the forward region has been shown to factorize~\cite{CFS}
as the convolution of a hard scattering kernel $H$ and two
non-perturbative objects obeying their own QCD evolution equations,
\begin{itemize}
\item a non-diagonal parton distribution~\cite{OFPD} $f(x,x-x_B,t,\mu)$,
\item the distribution amplitude of the produced meson $\phi(z,\mu)$,
\end{itemize}
as
\begin{eqnarray}
M=\int_0^1\! dz\,\int_0^1\! dx\, f(x,x-x_B,t,\mu)\nonumber\\
\times H(Q^2,x/x_B,z,\mu)\,\phi(z,\mu).
\label{CFS}
\end{eqnarray}

The introduction of generalized distribution amplitudes allows one to
enlarge the scope of this factorization property to the case of
non-resonant $\pi\pi$ production. This means that non-resonant $\pi\pi$
emission (including the $\pi^0\pi^0$ channel absent in $\rho$-decay)
will have the same scaling behavior as $\rho$-production. Its
amplitude is obtained in a straightforward way from~(\ref{CFS}) by
replacing $\phi \to \Phi$.

\section{CONCLUSION}

To our knowledge no data exist as yet for exclusive two-meson
production in $\gamma^* \gamma $-collisions in the kinematical domain
where QCD factorization would enable us to extract generalized
distribution amplitudes.  An estimation of expected yields at various
energies is under way~\cite{Us}.  Promising experimental set-ups
should be the high luminosity, medium energy $e^+ e^-$ colliders known
as the B-factories at SLAC and KEK. The hadronic physics potential of
these new machines is greatly enhanced by this new instance of
factorization.

\section*{Acknowledgments}

We gratefully acknowledge many discussions with O. Teryaev.  This work
has been partially funded through the European TMR Contracts
No.~FMRX-CT96-0008: Hadronic Physics with High Energy Electromagnetic
Probes and No.~FMRX-CT98-0194: Quantum Chromodynamics and the Deep
Structure of Elementary Particles. SUBATECH is Unit\'e mixte 6457 de
l'Universit\'e de Nantes, de l'Ecole des Mines de Nantes et de
l'IN2P3/CNRS. Centre de Physique Th\'eorique is Unit\'e mixte C7644 du
CNRS.


\begin{thebibliography}{99}
\bibitem{DGPT} M. Diehl, T. Gousset, B. Pire and O.V. Teryaev,
Phys.\ Rev.\ Lett.\ {\bf 81}, 1782 (1998).

\bibitem{MRG} D. M\"uller {\it et al.}, Fort. Phys. {\bf 42}, 101
(1994), hep-ph/9812448.

\bibitem{Fre} A. Freund, hep-ph 9903489.
  
\bibitem{RR} S. Ong, Phys.\ Rev.\ {\bf D52}, 3111 (1995);\\ R. Jakob,
  P. Kroll and M. Raulfs, J. Phys. {\bf G22}, 45 (1996);\\ P. Kroll
  and M. Raulfs, Phys.\ Lett.\ {\bf B387}, 848 (1996);\\ A.V.
  Radyushkin and R.T. Ruskov, Nucl.\ Phys.\ {\bf B481}, 625 (1996);\\ 
  I.V. Musatov and A.V. Radyushkin, Phys.\ Rev.\ {\bf D56}, 2713
  (1997).

\bibitem{pi-exp} CLEO Collab., J. Gronberg {\it et al.}, Phys.\ Rev.\ 
  {\bf D57}, 33 (1998).

\bibitem{OFPD} X. Ji, Phys.\ Rev.\ Lett.\ {\bf 78}, 610 (1997);\\ A.V.
  Radyushkin, Phys.\ Rev.\ {\bf D56}, 5524 (1997);\\ J.C. Collins and
  A.  Freund, Phys.\ Rev.\ {\bf D59}, 074009 (1999).

\bibitem{DGPR} M. Diehl {\it et al.},  Phys.\ Lett.\ {\bf B411}, 193
(1997).

\bibitem{KSZ} For a recent review and references, see M. Krawczyk,
M. Staszel and A. Zembrzuski, hep-ph/9806291.

\bibitem{Pol} M.V. Polyakov, hep-ph/9809483.

\bibitem{Poly-W} M.V. Polyakov and C. Weiss, hep-ph/9902451.

\bibitem{ERBL} G.P. Lepage and S.J. Brodsky, Phys.\ Lett.\ {\bf B87},
  359 (1979);\\ A.V. Efremov and A.V. Radyushkin, Phys.\ Lett.\ {\bf
    B94}, 245 (1980).
  
\bibitem{Chase} M.K. Chase, Nucl.\ Phys.\ {\bf B174}, 109 (1980).

\bibitem{Bud} V.M. Budnev {\it et al.}, Phys.\ Rept.\ {\bf C15}, 181
(1975).

\bibitem{Us} M. Diehl, T. Gousset and B. Pire, in preparation.
  
\bibitem{JLAB} M. Diehl, T. Gousset and B. Pire, to appear in the
  Procs.\ of the Workshop on Exclusive and Semiexclusive Processes at
  High Momentum Transfer, Jefferson Lab, USA, 20--22 May 1999.

\bibitem{CFS} J.C. Collins, L. Frankfurt and M. Strikman, Phys.\
Rev.\ {\bf D56}, 2982 (1997).

\end{thebibliography}
\end{document}